\documentclass[journal]{IEEEtran}
\IEEEoverridecommandlockouts
\usepackage{amsmath,amssymb,amsfonts}
\usepackage{algorithmic}
\usepackage{graphicx}
\usepackage{textcomp}
\usepackage{xcolor}
\usepackage{comment}
\usepackage{authblk}
\usepackage[font=footnotesize,skip=0pt]{caption}
\usepackage{multicol}
\usepackage{color}
\usepackage{graphicx}  % remove 'demo' option for your real document
\usepackage{soul}
\usepackage{physics}

\usepackage[ruled,vlined,linesnumbered]{algorithm2e}
\usepackage{comment}
\usepackage[numbers,sort&compress,square]{natbib}
\usepackage{enumerate}
\usepackage[numbers,sort&compress,square]{natbib}
\usepackage{pdfpages}

\usepackage[bookmarks=false, hidelinks]{hyperref}
\usepackage[normalem]{ulem}
\usepackage{float}
\usepackage{scalerel}
\usepackage{tikz}
\usetikzlibrary{svg.path}

\definecolor{orcidlogocol}{HTML}{A6CE39}
\tikzset{
  orcidlogo/.pic={
    \fill[orcidlogocol] svg{M256,128c0,70.7-57.3,128-128,128C57.3,256,0,198.7,0,128C0,57.3,57.3,0,128,0C198.7,0,256,57.3,256,128z};
    \fill[white] svg{M86.3,186.2H70.9V79.1h15.4v48.4V186.2z}
                 svg{M108.9,79.1h41.6c39.6,0,57,28.3,57,53.6c0,27.5-21.5,53.6-56.8,53.6h-41.8V79.1z M124.3,172.4h24.5c34.9,0,42.9-26.5,42.9-39.7c0-21.5-13.7-39.7-43.7-39.7h-23.7V172.4z}
                 svg{M88.7,56.8c0,5.5-4.5,10.1-10.1,10.1c-5.6,0-10.1-4.6-10.1-10.1c0-5.6,4.5-10.1,10.1-10.1C84.2,46.7,88.7,51.3,88.7,56.8z};
  }
}

\newcommand\orcidicon[1]{\href{https://orcid.org/#1}{\mbox{\scalerel*{
\begin{tikzpicture}[yscale=-1,transform shape]
\pic{orcidlogo};
\end{tikzpicture}
}{|}}}}

\def\BibTeX{{\rm B\kern-.05em{\sc i\kern-.025em b}\kern-.08em
    T\kern-.1667em\lower.7ex\hbox{E}\kern-.125emX}}
\begin{document}

\title{MMC-Based Distributed Maximum Power Point Tracking for Photovoltaic Systems}%*\\
\author{Farog Mohamed, Shailesh Wasti, Shahab Afshar, Pablo Macedo, and Vahid Disfani\\
ConnectSmart Research Laboratory, University of Tennessee at Chattanooga, TN 37403, USA \\
Emails: 
%Shahab-Afshar@mocs.utc.edu, Pablo-Macedo@mocs.utc.edu,
farog-mohamed@mocs.utc.edu, vahid-disfani@utc.edu

\thanks{\textcopyright 2020 IEEE.  Personal use of this material is permitted.  Permission from IEEE must be obtained for all other uses, in any current or future media, including reprinting/republishing this material for advertising or promotional purposes, creating new collective works, for resale or redistribution to servers or lists, or reuse of any copyrighted component of this work in other works.}}

\maketitle

\begin{abstract}
This paper proposes a novel topology for grid connected photovoltaic (PV) system based on modular multilevel converter (MMC). In this topology, a PV array is connected to capacitors of each submodule (SM) of the MMC through a DC-DC boost converter with maximum power point tracking (MPPT) control.  This topology will maximize the efficiency of the system in the case of partial shading conditions, as it can regulate the SM capacitor voltages independently from each other to realize distributed MPPT. A model predictive control is used to track the AC output current, balance the SMs capacitor voltages, and to mitigate the circulating current. The proposed PV generation topology with 7 level MMC system validity has been verified by simulations via MATLAB/Simulink toolbox under normal operation, partial shading and PV array failure.
\end{abstract}

\begin{IEEEkeywords}
 Maximum power point tracking, modular multilevel converter, model predictive control, partial shading, and photovoltaic system.
\end{IEEEkeywords}

\section{Introduction}

Solar PV energy experienced tremendous growth over the last years,  Solar PV represented about 47\% of newly installed renewable power capacity in 2016 \cite{secretariat2017renewables}. The main reason for this growth is the continuous drop in PV modules cost which eventually led to the development of large scale PV power plants.  The main drawbacks of solar PV energy are intermittency, uncertainty and its variability with time and location, which necessitate advanced solutions to capture maximum solar energy at any time. 

There are numerous power electronic solutions to address these issues. Most of the PV connection topologies have been focused on small or medium-scale systems with string and multi-string inverter configurations. However, the conventional technology practiced so far for utility-scale PV power plants is the centralized topology. In this topology, PV cells are connected in series to form a module, and modules are connected in series and parallel to form array of strings. The power from the array is fed to a single string inverter controlled by an MPPT algorithm to capture the maximum solar energy available \cite{deline2011performance}. However, this topology leads to performance loss due to shading losses and mismatch between PV modules.
{Module-level power electronics (MLPE)--also known as distributed MPPT (DMPPT)--is a newly proposed topology which decouples the maximum power point of individual modules from overall MPPT of PV system by introducing separate DC-DC converters for each module. This topology has been studied extensively in the literature and it was proven to be effective as it can recover 30\%–40\% of this power loss due to partial shading \cite{deline2011performance, hanson2014partial}.}

With the recent advancement in smart grid, utility-scale PV plants and multilevel technologies, converter topologies have to be more flexible, reliable with modular structure, and multilevel voltage \cite{xue2011towards}. MMC is an advanced power electronics converter that offers these characteristics \cite{disfani2015fast,qin2012predictive}. With a wide range of potential applications in the medium and high-power applications, most of MMC application has been limited to high-voltage direct current (HVDC)  \cite{qin2012predictive,khanal2019novel,khanal2019reduced}. There are a few research works that employ MMC for integration of wind farms \cite{gnanarathna2010modular} and photovoltaic systems \cite{ nademi2016comparative, mei2013modular, rajasekar2012solar, echeverria2013multi, alajmi2011modular, ramya2015design, ramya2015switching, rivera2013modular,rong2016novel, khazaei2019novel}.

In \cite{nademi2016comparative, mei2013modular, rajasekar2012solar}, an MMC solution is proposed to integrate series-connected PV arrays to power grids, where the PV system is connected to the DC link of MMC through a DC/DC converter. This solution does not address the partial shading problem. \cite{echeverria2013multi} proposes an MMC based HVDC system where the PV system is connected to the MMC DC link through a two-stage DC-DC converter. In \cite{alajmi2011modular},  a single phase MMC solution connected to a DC-DC converter with MPPT control is proposed to interface the PV system, but the partial shading problems persist using this solution.  \cite{ramya2015design} studies PV array connection to the DC side of the MMC directly but the main focus is on the output filter design. \cite{ramya2015switching} studies the switching loss and total harmonic distortion (THD) analysis of MMC in a grid connected PV systems with different Switching Frequency.

An MMC-based MLPE solution is proposed in \cite{rivera2013modular, rong2016novel} where each SMs is connected to one PV module. These solutions are however very complicated for implementation and lack circulating current control as one of the main objectives in MMC control design. To address the circulating current issue, \cite{khazaei2019novel} proposes a similar MMC-based MLPE solution with circulating current control. The solution, however, suffers from SM voltage balancing perspective. In PV interconnection to power grid where no energy storage systems are available, it is required that there be no mismatch between the energy captured from PV modules and the energy transferred to the grid. Any power mismatch is stored in or provided by the MMC SM capacitors can cause long-term deviation in SM capacitor voltages. Thus, it is physically impossible to capture maximum solar power and perform demand response at the same time unless energy stored in capacitors is affected. The paper also fails to provide any SM capacitor voltage results to prove otherwise. Furthermore, no research work has studied MMC operation in the event of PV array failure.
%\Farog{ }

To address these challenges, this paper proposes a novel integrated power electronics system that utilizes MMC to connect distributed PV systems using DMPPT across the MMC SMs, given the benefits of DMPPT in solving the partial shading performance issues. An average SM capacitor voltage control algorithm is  proposed in this paper to ensure all the captured solar energy is transferred to the AC grid. The model predictive control (MPC) strategy proposed in \cite{disfani2015fast} is implemented to get the best switching sequences of SMs to control ac-side current, capacitor voltage, and circulating current simultaneously. The proposed solutions are scalable from residential systems up to utility scale systems, by adjusting power and voltage ratings of MMC components or by increasing the number of SMs on each arm. The algorithms are tested against different case studies to demonstrate their performance.

The rest of the paper is organized as follows. Section \ref{sec:topology} presents the proposed system topology. Section \ref{sec:Control Strategy} presents the Proposed Novel Control Strategy for the MMC and SM voltage control. Section \ref{sec:Simulation} reports three case studies simulation analysis. Section \ref{sec:conclusion} concludes the paper.

\section{The proposed topology of the MMC with distributed PV system} 
\label{sec:topology}

\subsection{System Topology}
\begin{figure}
\centering
\includegraphics[width=0.45\textwidth, height=0.25\textheight]{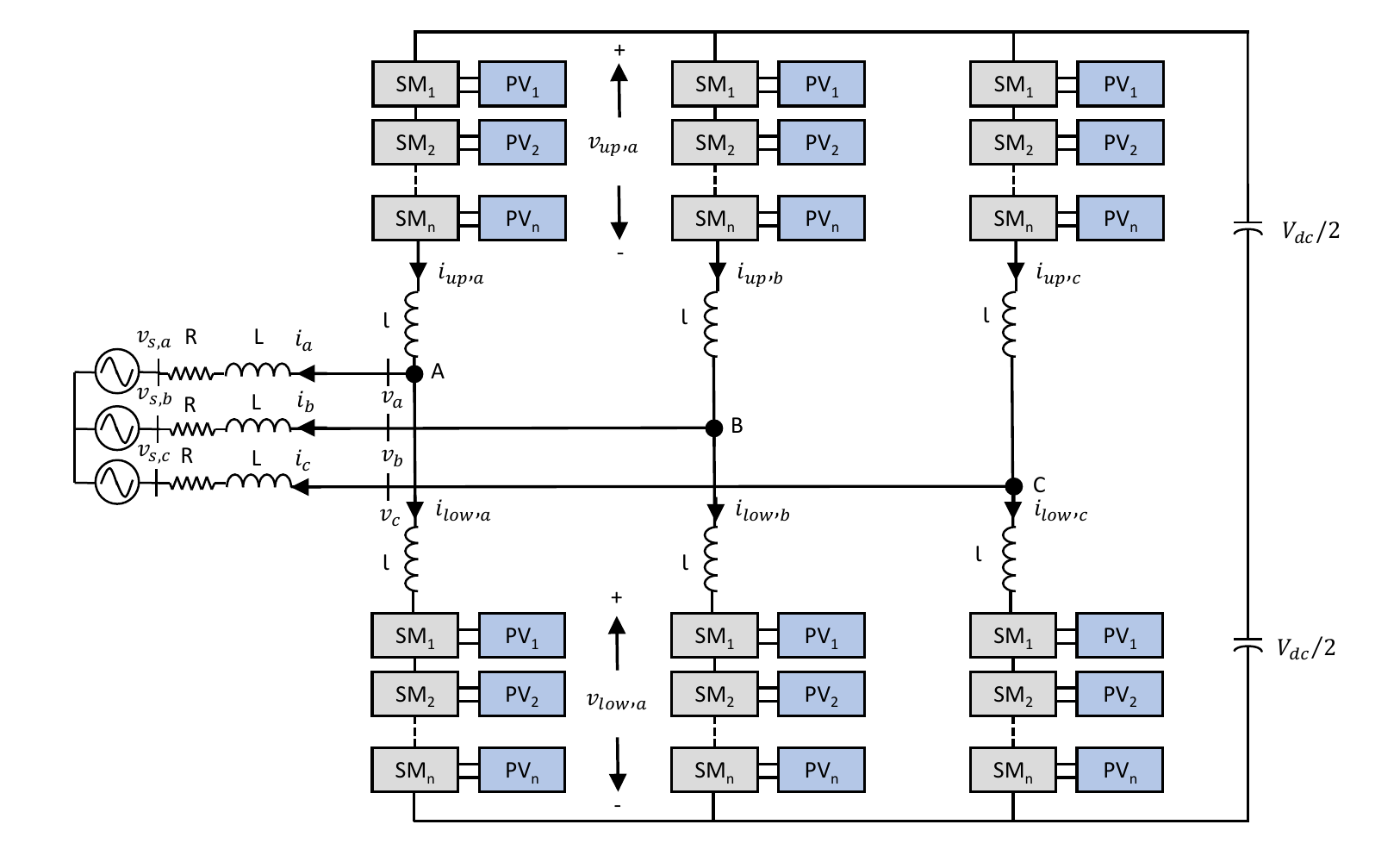}
\caption{Topology of the proposed MMC-PV system  \cite{khazaei2019novel}}
\label{fig:MMC_topolgy}    
\end{figure}

The proposed three-phase MMC solution is shown in Fig.~\ref{fig:MMC_topolgy}. It consists of 2 arms at each phase, where each arm has $n$ SMs. The SMs are half bridge SMs with two IGBT switches and a capacitor, each connected to one PV module through a DC optimizer. Each SM voltage is either zero or its DC link voltage $v_ci$ depending on the SM switches states. The MMC is connected to the three-phase AC system at the point of common coupling (A, B, and C) through a filter with resistance and inductance of $R$ and $L$ on each phase. Each arm has two inductors ($l$) placed for current control and faults limiting.\\

\subsection{MMC Discrete Model }\label{MMC-Maths}
The discrete model of MMC used in this paper was proposed in \cite{disfani2015fast}, where Euler’s approximation of the current derivative that represents the next step value for the AC-side current can be expressed as:
\begin{align}
&\begin{matrix}i(t+T_s)=\frac{1}{K'}{\left(  {\frac {v_{low}(t+T_s)-v_{up}(t+T_s)}{2} -v_s(t+T_s)+\frac{L'}{T_s}i(t)}\right)}\end{matrix}
\label{Idis}
\end{align}

\noindent where the time step $T_s$ is small sampling time, $L'=L+l/2$ and $K'=R+L'/T_s$. The measured values at the current time are denoted by time indices $(t)$ and the predicted values for the next time step are denoted by  $(t+T_s)$. The sampling frequency is assumed to be significantly higher compared to the grid frequency, the predicted value of grid voltage $v_s(t+T_s)$ can be replaced by its measured value $v_s(t)$. The predicted capacitor voltage of individual SMs on upper-level and lower-level arms is equal to: 
\begin{align}
&v_{Cj}(t+T_s)=v_{Cj}(t)+\left(\frac{T_s i_{up}(t)}{C}\right)u_j(t+T_s) && \forall_{j\in [1,n]}\label{VCupdis} \\
&v_{Cj}(t+T_s)=v_{Cj}(t)+\left(\frac{T_s i_{low}(t)}{C}\right)u_j(t+T_s) &&\forall_{j\in [n+1,2n]} \label{VClowdis}
\end{align}
Where $u_j(t+T_s)$ is the status of $j$-th SM.
Thus, the predicted voltages across upper-level and lower-level arms and circulating current for the next step are defined as:
\begin{align}
&v_{up}(t+T_s)=\sum_{j=1}^{n} v_{Cj}(t+T_s)u_j(t+T_s)  \label{Vupdis} \\
&v_{low}(t+T_s)=\sum_{j=n+1}^{2n} v_{Cj}(t+T_s)u_j(t+T_s) \label{Vlowdis} \\
&i_z(t+T_s)=\frac{T_s}{2l}\left(V_{DC}-v_{low}(t+T_s)-v_{up}(t+T_s)\right)+i_z(t)
\label{Izdis}
\end{align}
\subsection{Distributed PV Modules}\label{MLPE}
\begin{figure}[h]
\centering
\includegraphics[width=0.45\textwidth , height=0.18\textheight]{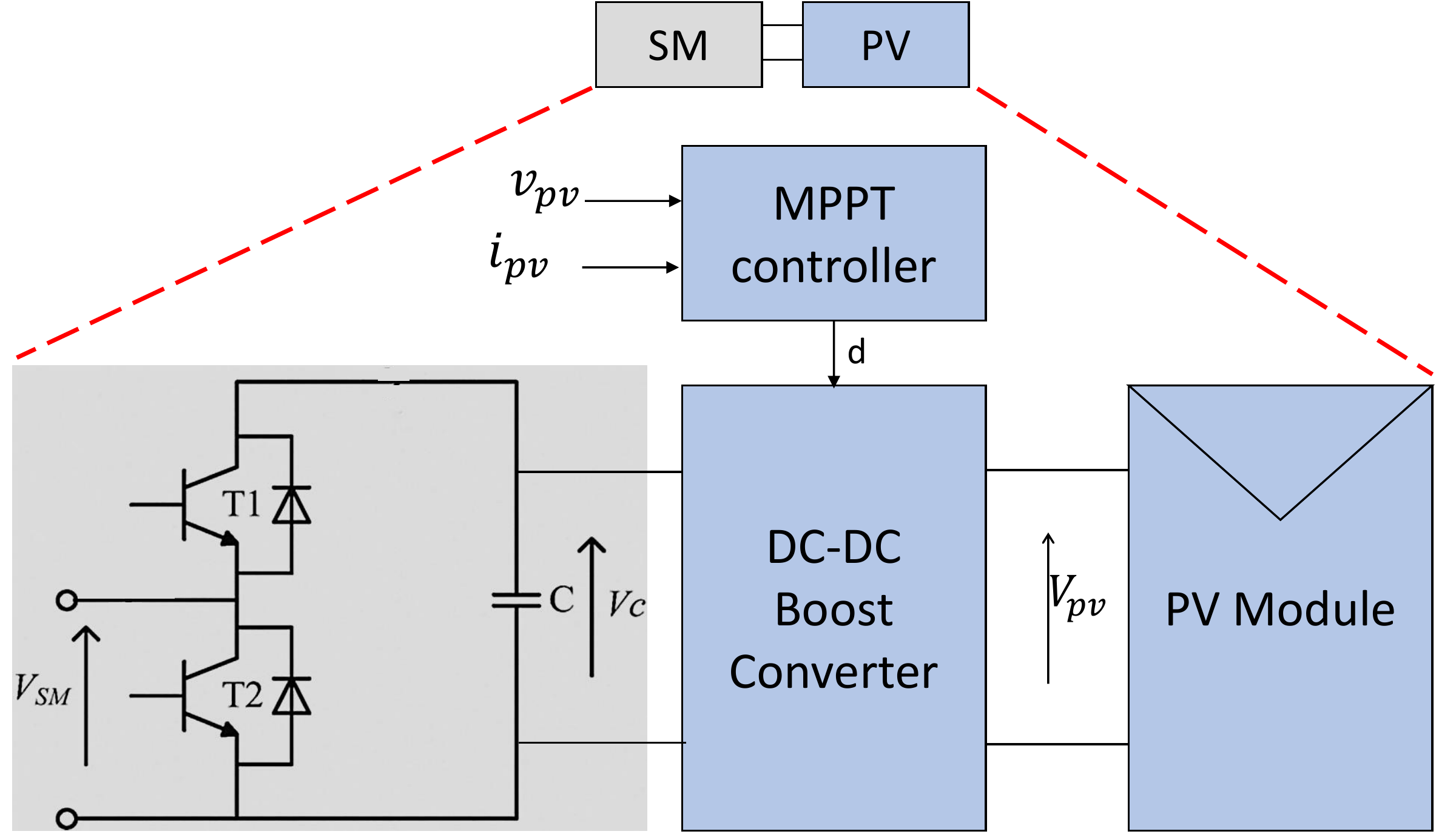}
\caption{SM and PV module connection.}
\label{fig:MLPE}    
\end{figure}

The proposed MMC-PV topology removes the PV modules strings connected to the MMC DC side and connect each PV module to SM DC link through a DC/DC converter with MPPT controller as shown in Fig.~\ref{fig:MLPE}. With this topology, in case of partial shading, the MPPT controller of each SM captures the maximum power of its PV module by regulating the voltage across the PV module on the MMP voltage $(V_{mmp})$ at any time. Different MPPT algorithms have been developed  \cite{de2012evaluation}. Among them, perturb and observe ($P\&O$) is one of the most common algorithms, where voltage control signal at each time step is defined based on the effect of the previous adjustment on the PV power output. See \cite{abdelsalam2011high, subudhi2012comparative} for more details.

\section{Proposed Control Strategy} 
\label{sec:Control Strategy}
\subsection{Model Predictive Control}
To effectively control the MMC, the optimal switching sequence is obtained by using the model predictive control (MPC) strategy from \cite{disfani2015fast}, which seeks the following objectives:
\begin{enumerate}[i.]
\item to track the ac-side current (i) of all phases to their reference values ($i_{ref}$),
\item to regulate all the submodules capacitor voltages to their nominal value ($V_{DC}/n$), and
\item to mitigate the circulating current ($i_z$)  between the converter phase legs.
\end{enumerate}
Assuming that the ideal value of corresponding variable for the next time step is donated by  $(\cdot)^*(t+T_s)$, the ideal values implying exact AC current tracking and exact circulating current suppression which is can be represented by $i(t+T_s)=i_{ref}$ and $i_z(t+T_s)=0$ respectively. The anticipated values of upper and lower level voltages of MMC are calculated as:

\begin{align}
&v_{up}^*=\left(\frac{V_{DC}}{2}+\frac{l}{T_s}i_z(t)\right)-
\left(K'i_{ref}+v_s(t)-\frac{L'}{T_s}i(t)\right)
\label{Vup*}\\
&v_{low}^*=\left(\frac{V_{DC}}{2}+\frac{l}{T_s}i_z(t)\right)+
\left(K'i_{ref}+v_s(t)-\frac{L'}{T_s}i(t)\right)
\label{Vlow*}
\end{align}

Let $\Delta i=i-i_{ref}(t+T_s)$, $\Delta v_{low}=v_{low}^*-v_{low}$, and $\Delta v_{up}=v_{up}^*-v_{up}$ donates the deviation of the corresponding variables from their ideal values. The deviation of the AC current and the circulating current from their ideal values are derived as:
\begin{align}
&\Delta i= \frac{1}{2K'}\left(\Delta v_{low}(t+T_s)-\Delta v_{up}(t+T_s)\right)
\label{Idis_error}\\
&i_z(t+T_s)=\frac{T_s}{2l}\left(\Delta v_{low}(t+T_s)+\Delta v_{up}(t+T_s)\right)
\label{Izdis_error}
\end{align}

Applying a weighted sum method to the optimization problem, the AC current  tracking and circulating current mitigation objectives with weights $w$ and $w_z$  respectively. The following multi-objective optimization problem describes the switching algorithm:

\begin{align}
&\min_U&& {\sum_{j=1}^{2n}\abs{v_{C_j}(t+T_s)-v_{C_j}(t)}}  \label{sorting_volt_obj}\\
&\min_U&&f=\left\{\begin{matrix}\frac{w}{2K'}\left|\Delta v_{low}(t+T_s)-\Delta v_{up}(t+T_s)\right|+\\
\\
\frac{w_z T_s}{2l}\left|\Delta v_{low}(t+T_s)+\Delta v_{up}(t+T_s)\right|\end{matrix}\right\}\label{selection_obj}\\
&\text{subject to:} &&~~\eqref{Idis}-\eqref{Izdis}\nonumber\\
&&& ~~U=[u_1,u_2,...,u_{2n}] : u_j \in \{0,1\}~~~~\forall_{j\in[1,2n]}\label{constraint_status_binary}
\end{align}

Where the first objective \eqref{sorting_volt_obj} regulates SM capacitor voltages and the second objective  \eqref{selection_obj} follows the reference values of AC current and circulating currents.
\subsubsection{SM Sorting} In this step, the SM capacitor voltage regulation objective function \eqref{sorting_volt_obj} is solved by sorting SMs effectively where the highest priority is given to the SMs contributing the most in voltage balancing. It starts by sorting the upper and lower arms SMs based on their expected capacitor voltages. Since these SM voltages increase or decrease based on the direction of $i_{up}$, the SMs are sorted based on their capacitor voltages in the descending order if $i_{up}<0$ or in the ascending order if $i_{up}\ge 0$. After sorting, define $V_{C_{up}}^{sort}=[V_{C_1}^{sort},...,V_{C_n}^{sort}]$ and $V_{C_{low}}^{sort}=[V_{C_{n+1}}^{sort},...,V_{C_{2n}}^{sort}]$ denote SM voltages on upper and lower arms respectively.

\subsubsection{SM selection} This step calculates the cumulative sum vectors of the components of $V_{C_{up}}^{sort}$ and $V_{C_{low}}^{sort}$ to get $V_{C_{up}}^{sum}$ and $V_{C_{low}}^{sum}$ as defined as below.
\begin{align}
&V_{C_{up}}^{sum}=\{\alpha_k:k=0,1,...,n\} \label{sum_up}\\
&V_{C_{low}}^{sum}=\{\beta_k: k=0,1,...,n\}  \label{sum_low}
\end{align}
where
\begin{align}
&\alpha_0=\beta_0=0\nonumber\\
&\alpha_k=\Sigma_{i=1}^{k}V_{C_i}^{sort}&\forall_{k\in [1,n]}\nonumber\\
&\beta_k=\Sigma_{i=n+1}^{n+k}V_{C_i}^{sort}&\forall_{k\in [1,n]}\nonumber
\end{align}

To minimize the objective function \eqref{selection_obj} the switching algorithm defines what combination of $(\alpha,\beta)$ is needed. In \cite{disfani2015fast}, it is proven that if $v_{up}^*\in[\alpha_i,\alpha_{i+1})$ and $v_{low}^*\in[\beta_j,\beta_{j+1})$, the optimal solution belongs to the set $\{(\alpha_i,\beta_j),(\alpha_{i+1},\beta_j), (\alpha_i,\beta_{j+1}),(\alpha_{i+1},\beta_{j+1})\}$. It means that it suffices to check the objective function for just 4 points instead of $n^2$ solutions to select the best SMs to switch on.

\subsubsection{Computation Requirements}
Computation expense to solve the sorting and selection process is one of the concerns of MMC switching through MPC. The switching frequency must be selected such that the computations at any time should take less than one-time step. If occasionally, the computation process takes more than a one time step at some point, the previous switching sequence is applied to the submodules.\\
\subsection{AC output Current Control}
\label{AC_control}
The reference AC waveforms $i_{abc}^{ref}$ are controlled to regulate the average voltage of SM capacitors on their nominal values to ensure that no extra energy is stored in the SM capacitors. This control is shown in Fig. \ref{fig:AC_control_circuit} below, it takes the average of measured SM capacitor voltages, the reference value of the average of SM capacitor voltages and the grid reference current to get the AC output current that should be sent to the grid. The PI controller acts upon the average SM capacitor voltage deviation from its nominal value to generate a control signal in terms of a reference current on a direct axis ($i_{d}$). With an addition of a quadrature axis component of current ($i_q$), the generated $dq$-component of current ($i_{dq}$) is added to the $dq$-components of actual output current ($i_{abc}$). The total output signal is finally converted into $abc$-frame, and the AC reference current ($i_{abc}^{ref}$) is fed to the MMC switching algorithm.
\begin{figure}[H]
\centering
\includegraphics[width=0.45\textwidth]{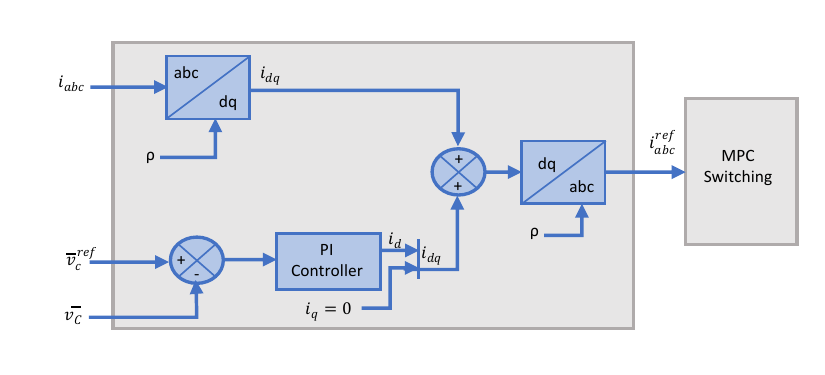}
\caption{AC output Current Control.}
\label{fig:AC_control_circuit}    
\end{figure}

\section{Case Study} 
\label{sec:Simulation}
\subsection{Simulation Setup}
The proposed topology is simulated on MATLAB and tested to verify the performance of MMC and the control methods. Since each SM is connected to a PV array, the focus is to control the SM capacitor voltage under partial shading conditions and guarantee the tracking of the maximum power. The system has 36 solar PV Panels individually controlled via $P\&O$ DMPPT method, and their Parameters are listed in Table \ref{PV_Data}. The MMC parameters are given in Table \ref{MMC_Parameter}. The PV array temperature input is assumed to be 25 ${C}^{\circ}$ all the time. To study the partial shading, the irradiance data were taken from the PV Power Research Plant of Tampere University \cite{Jussi:2018}. The simulations were run for 3 seconds and the system was tested for three case studies, normal operation, partially cloudy and PV module failure. Given that the MMC simulated has $6$ SMs, the case studies are designed such that no partial shading occurs on SMs 1-4 of all arms; thus, they are exposed to 100\% of their associated irradiance. On the contrary, SMs 5-6 of all arms are shaded and receive 20\% of their associated irradiance. To simulate PV module failure, the PV modules connected to SM 1 of all arms fail at $t=2s$ and remain disconnected till the end of simulation time. The following discussions are focused on performance of MMC to realize DMPPT.\\

\begin{table}[htbp]
\caption{Solar PV Array Data} %title of the table
\centering % centering table
\begin{tabular}{|c|c|} % creating 2 columns
\hline %inserting line
{Parameter} & {value} \\ [0.5ex]
\hline % inserts single-line
Module & SunPower SPR-305E-WHT-D\\ % Entering row contents
Maximum Power & 305.226 W\\
Cells per module & 96\\
Open circuit voltage (${V}_{oc}$) & 64.2 V\\
Short-circuit current (${I}_{sc}$) & 5.96 A\\
Voltage at MPP (${V}_{MPP}$) & 54.7 V\\
Current at MPP (${I}_{MPP}$) & 5.58 A\\
Temperature coefficient of ${V}_{oc}$ & -0.27269 ${\%}/{C}^{\circ}$ \\
Temperature coefficient of ${I}_{sc}$ & 0.061745 ${\%}/{C}^{\circ}$ \\ [1ex] % [1ex] adds vertical space
\hline % inserts single-line
\end{tabular}
\label{PV_Data}
\end{table}

\begin{table}[htbp]
\centering
\caption{MMC Parameters}
\begin{tabular}{|c|c|}
  \hline
  Parameter&Value\\
  \hline
  Number of submodules per arm & 6\\
  Active power delivery & $10.9$ kW\\
  Nominal DC voltage $V_{DC}$ & $600$ V\\
  Sampling period $T_s$&$25$ $\mu$s\\
  Output current reference $I_{ref}$ & $16$ A\\
  Submodule capacitor $C_{sm}$ & $5000$ $\mu$F\\
	$R$&$0.003$ $\Omega$\\
	$L$&$5$ mH\\
	$l$&$5$ mH\\
 \hline
	\end{tabular}
\label{MMC_Parameter}
\end{table}
\subsection{MMC Control Performance Under Partial Shading Condition}
In this case, PV arrays on the MMC SMs receive fluctuating irradiance in normal operation. Fig.~\ref{fig:Irr_P} shows the irradiance and the output power of phase A upper arm SMs, It shows that the power output of PV modules change with their own irradiance. The PV arrays on SM 5-6 receive extremely low irradiance (20\%) due to partial shading. From Fig.~\ref{fig:Irr_P}, power output of the PV systems connected to these SMs is around 100 Watt since the irradiance never exceeds 250 $Watt/{m}^{2}$. Although these PV modules are partially shaded, the PV modules connected to SMs 1-4 are not affected and work at their maximum power point. Therefore, the efficiency of the system is maximized by individually controlling the PV arrays. 

\subsection{MMC Control Performance Under PV Modules Failure and Partial Shading Condition}
\begin{figure}[h]
\centering
\includegraphics[width=0.5\textwidth]{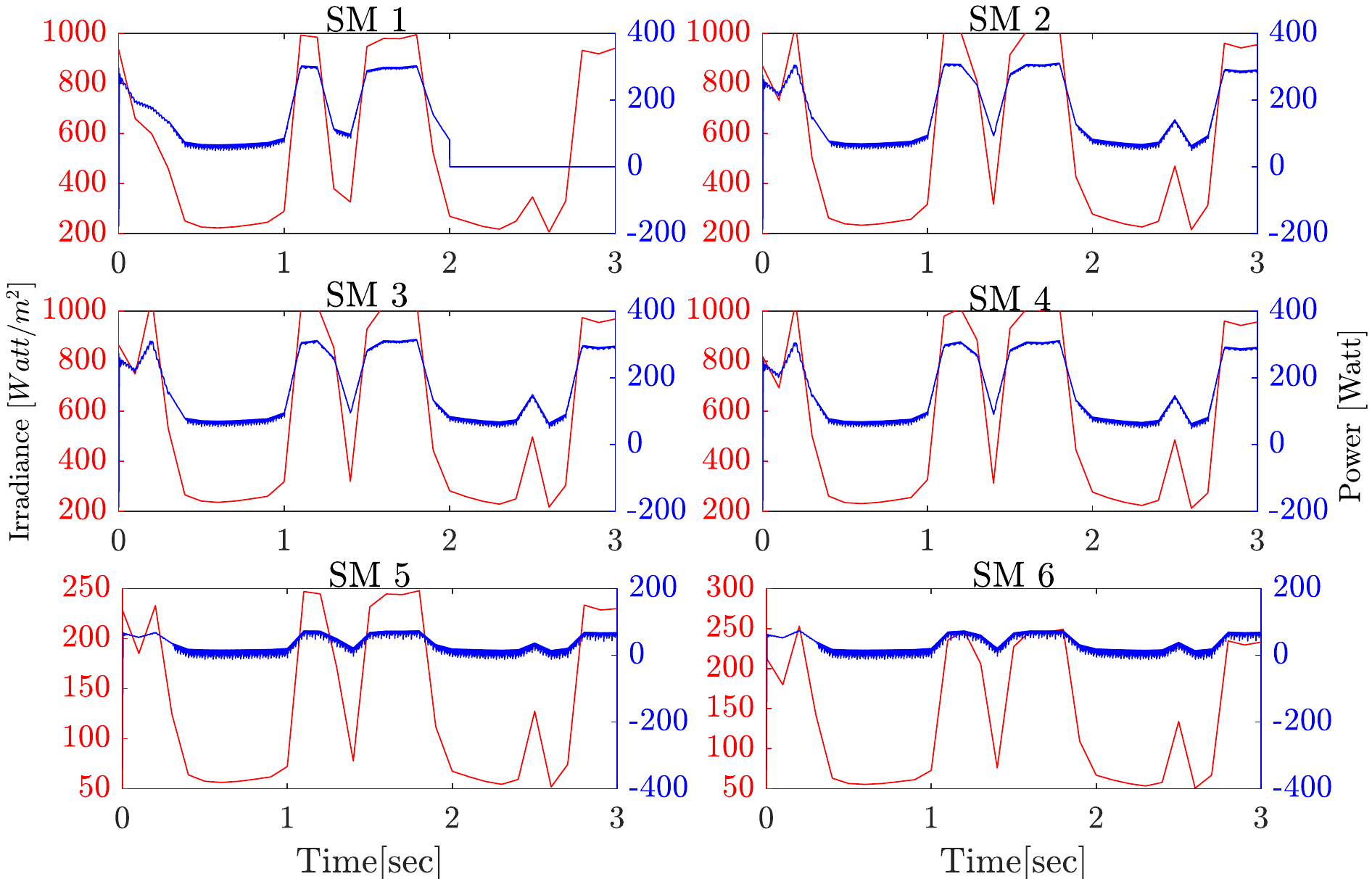}
\caption{Phase A upper arm submodules irradiance and power. The figure is shown in double-axis format where irradiance and power are shown in different colors (red and blue) on the left and right axes, respectively.}
\label{fig:Irr_P}    
\end{figure}

Between $t = 2s$ and $t = 3s$, the PV module connected to SM 1 of each arm fails and is disconnected from the SM. As illustrated in Fig.~\ref{fig:Irr_P}, the power output of the PV module of SM 1 power drops to 0 in second 2. Meanwhile, the other SMs work as normal proving the fact that neither partial shading nor the failure of any PV module affects the DMPPT of the others using the MMC solution proposed in this paper. 

\subsection{MMC Modulation (MPC) Performance}
Other than capturing the maximum power from individual PV modules, it is important to ensure that MMC operates as expected at all times. Parameters of interest are SM capacitor voltages, AC current waveform, and circulating current.  

\begin{figure}[h]
\centering
\includegraphics[width=0.5\textwidth]{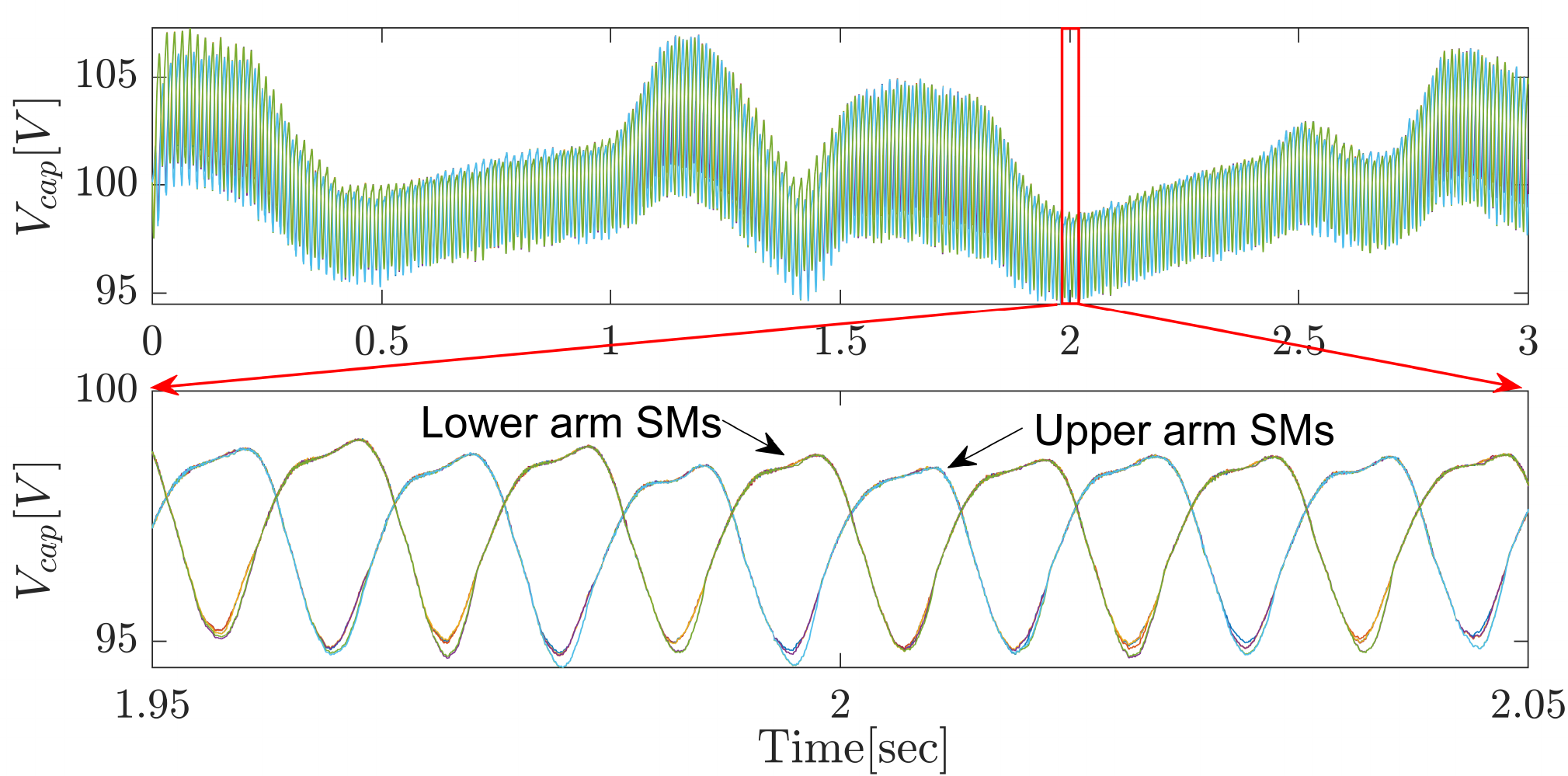}
        \caption{Phase A upper and lower SMs voltage.}
\label{fig:Vcap}    
\end{figure}

Fig.~\ref{fig:Vcap} shows capacitor voltages of all SMs on phase A upper and lower arms. The results show that capacitor voltage waveforms of upper arm SMs match each other at any time regardless of partial shading or PV failure. The same behavior is observed for capacitor voltage waveforms of lower arm SMs. This proves perfect performance of the SM capacitor voltage balancing implemented in MPC-based switching algorithm. Moreover, the fact that the average SM capacitor voltage of MMC is controlled within a $\pm5\%$ demonstrates that the AC output current control shown in Fig. \ref{fig:AC_control_circuit} successfully transfers the entire solar energy captured from PV modules to the AC power grid and no energy is accumulated on SMs of MMC. 

\begin{figure}[H]
\centering
%\hspace*{-3mm} 
\includegraphics[width=0.5\textwidth]{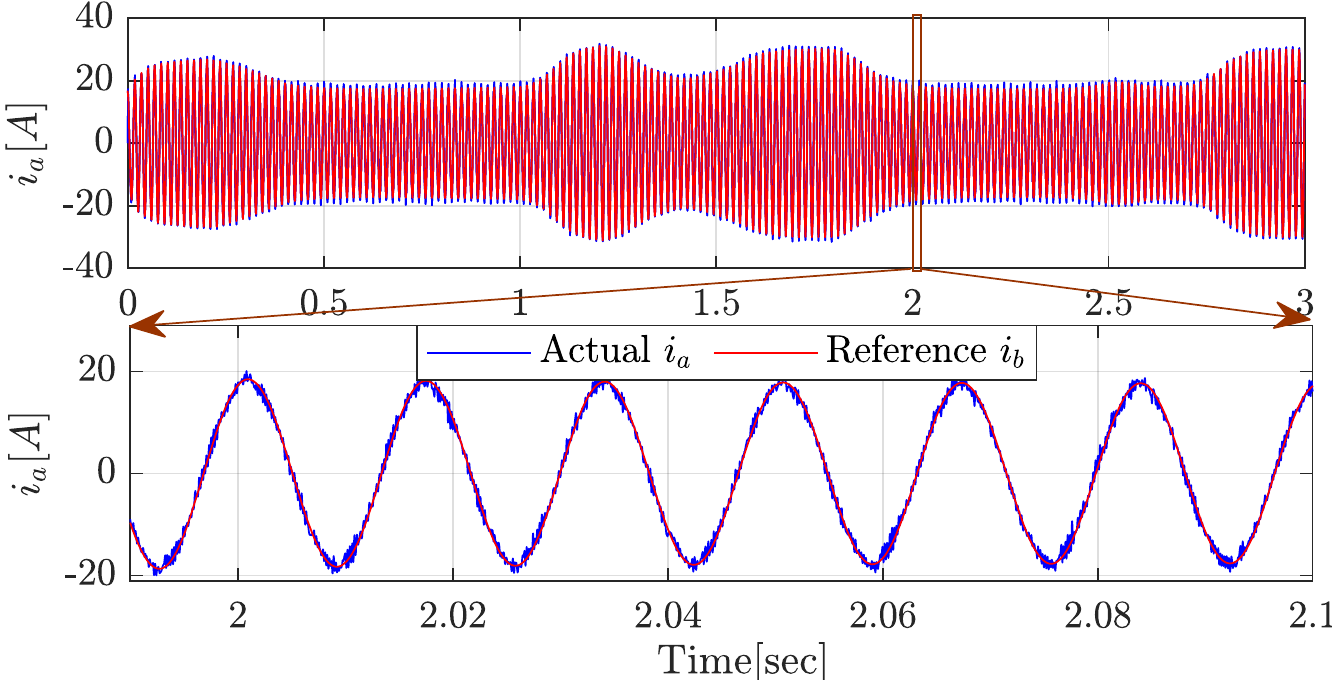}
\caption{AC output current tracking.}
\label{fig:AC_C}    
\end{figure}

\begin{figure}[h]
\centering
\includegraphics [width=0.5\textwidth]{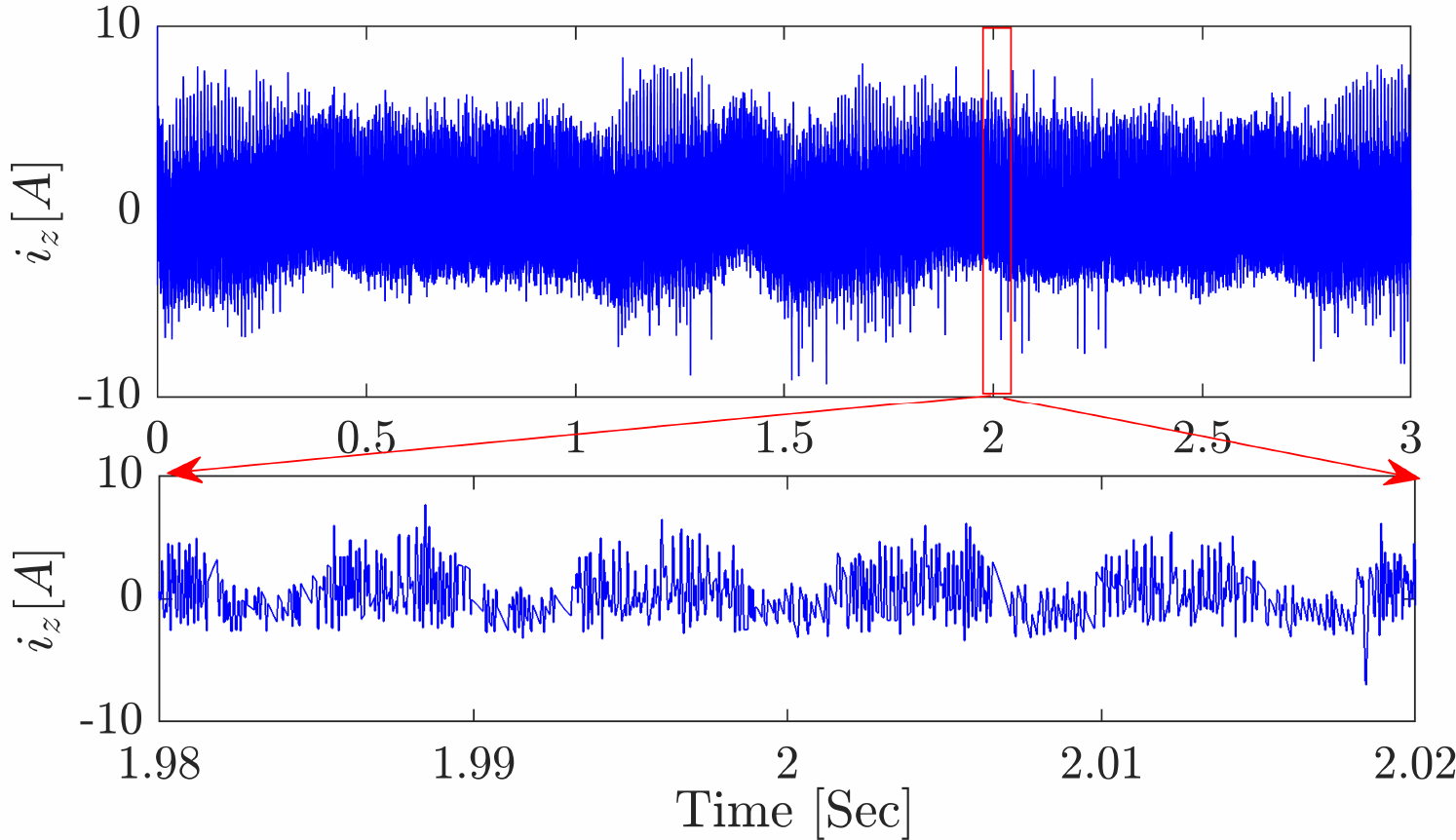}
\caption{Phase A circulating Current.}
\label{fig:Cirr_C}    
\end{figure}

The other objectives of the MPC-based switching algorithm are AC current waveform tracking and circulation current mitigation. Despite all the abnormal PV conditions simulated, Fig.~\ref{fig:AC_C} illustrates that the AC output current of phase A is tracked perfectly during the whole simulation time. Circulating current of phase A is also minimized around zero throughout the simulations as depicted in Fig.~\ref{fig:Cirr_C}. Similar results are observed for other phases of MMC but not shown here to avoid the repetition of similar results.

\subsection{MPC Computation Time}
In order to solve the MPC problem through the sorting and selection algorithms, we dedicated a separate computer with 3.60 GHz CPU and 64.0 GB RAM. The MPC computation takes less than one time step in 99.95\% of the time.\\

\section{Conclusion} 
\label{sec:conclusion}
An MMC-based solution to interconnect PV systems through DMPPT is proposed in this paper. The proposed power electronic solution includes single PV modules connected to SMs of MMC through DC-DC converters to realize MLPE. An MPC-based switching algorithm is employed in this paper. The power exchange between MMC and the AC power grid is also controlled to maintain the average SM capacitor voltages of the MMC as a proxy of the energy stored inside MMC. The results demonstrate that at any abnormal condition including partial shading and PV failure, the proposed solution leads to capturing maximum solar energy and perfect SM capacitor voltage balancing, AC current tracking, and circulating current suppression. Future work will focus on integration of various distributed energy resources to the grid using MMC.

\bibliographystyle{IEEEtran}
{\footnotesize \bibliography{IEEEabrv,MMC}}
\end{document}